\documentclass[12pt]{article}
\usepackage{graphicx} 
\usepackage[margin=1in]{geometry}
\usepackage{amsmath,bm,amssymb,amsthm}
\usepackage{natbib}
\usepackage{setspace}
\usepackage{longtable}
\newcommand{\indep}{\raisebox{0.05em}{\rotatebox[origin=c]{90}{$\models$}}}
\usepackage{enumerate}
\usepackage{enumitem}  
\usepackage{hyperref}
\usepackage{multirow}
\usepackage{lscape}
\usepackage{longtable}
\usepackage{lipsum} 
\usepackage{tikz}
\usetikzlibrary{positioning,tikzmark,fit}
\hypersetup{
    colorlinks=true,
    linkcolor=red,
    citecolor=blue,
    filecolor=magenta,      
    urlcolor=cyan,
    pdfpagemode=FullScreen,
    }
\usepackage{textcomp}
\usepackage{algorithm}
\usepackage{algpseudocode}
\usepackage[english]{babel}
\newtheorem{theorem}{Theorem}
\newtheorem{lemma}[theorem]{Lemma}

\usepackage{setspace}
\newcommand{\tred}[1]{\textcolor{red}{#1}}

\title{Estimate Time-Varying Exposure Effects via Ensemble Learning-based Marginal Structural Model with Application to Adolescent Cognitive Development Study}
 \author
{Zhiwei Zhao\\
{zzhao94@umd.edu} \\
\small{Department of Mathematics, University of Maryland,
College Park}\\
Chixiang Chen, Shuo Chen \\
{Chixiang.Chen@som.umaryland.edu,  ShuoChen@som.umaryland.edu}\\
\small{Division of Biostatistics and Bioinformatics,
University of Maryland School of Medicine} \\
}

\date{}
\providecommand{\keywords}[1]{\textbf{\textit{Keywords:}} #1}

\begin{document}
\doublespacing
\maketitle

\begin{abstract}
Evaluating the effects of time-varying exposures is essential for longitudinal studies. The effect estimation becomes increasingly challenging when dealing with hundreds of time-dependent confounders. We propose a {\textit{Marginal Structure Ensemble Learning Model} (\textbf{MASE})} to provide a marginal structure model (MSM)-based robust estimator under the longitudinal setting. The proposed model integrates multiple machine learning algorithms to model propensity scores and a sequence of conditional outcome means such that it becomes less sensitive to model mis-specification due to any single algorithm and allows many confounders with potential non-linear confounding effects to reduce the risk of inconsistent estimation. Extensive simulation analysis demonstrates the superiority of  \textbf{MASE} over benchmark methods (e.g., MSM, G-computation, Targeted maximum likelihood), yielding smaller estimation bias and improved inference accuracy. We apply \textbf{MASE} to the adolescent cognitive development study to investigate the time-varying effects of sleep insufficiency on cognitive performance. The results reveal an aggregated negative impact of insufficient sleep on cognitive development among youth.

\end{abstract}


\section{Introduction}

Recent large-scale longitudinal observational studies have collected extensive data, including time-varying exposures, multiple longitudinal outcomes, and hundreds of important time-dependent covariates. For instance, our current research is inspired by an investigation into the influence of insufficient sleep on adolescent cognitive development, utilizing data from the ABCD study \citep{yang2022effects}. The ABCD study collects data from $11,868$ adolescent participants at ages $10$, $12$, and $14$. During each visit, various measures are recorded, including sleep patterns, cognitive performance, and a range of psychographic variables. Insufficient sleep is a modifiable and health-related factor that may adversely affect the cognitive development of young individuals \citep{de2017effects}. A previous cross-sectional study revealed association between insufficient sleep and resultant differences in brain functions and structures \citep{dutil2018influence}. Recent studies researched on the longitudinal ABCD data and revealed the relationship between insufficient sleep and psychiatric problem \citep{cheng2021sleep}, and neurocognitive development \citep{yang2022effects}. However, those findings are mostly based on association-based analysis, which lacks causal interpretation, and did not consider time-varying trajectory effects. Therefore, we aim to re-evaluate the longitudinal (potentially causal) impact of insufficient sleep on cognitive outcomes among adolescent participants. The findings from this study can provide critical evidence to address sleep insufficiency, such as promoting strategies to reduce prolonged use of electronic devices as individuals age, ultimately helping to mitigate the risk of cognitive decline in young individuals. To consistently study the impact of insufficient sleep over the time using an observational data, it is important to control time-varying covariates (e.g., environmental factors, nutrition, and socioeconomic status) serving as potential confounders. Analyzing such a data requires advanced causal inference tools.


The current literature offers methods to address limited number of confounders with linear confounding effects, including inverse probability weighting \citep{horvitz1952generalization, robins1994estimation}, conditional mean imputation \citep{robins1999testing}, and matching \citep{rubin1973matching}. 
To estimate the effects of time-varying exposures in the presence of time-varying confounders, the Marginal Structural Model (MSM, \citealp{robins2000marginal,robins2000marginalb,babino2019multiple}) has become a popular tool in longitudinal causal inference. This method constructs a pseudo-population and incorporates both confounders and time-varying exposures into weights. However, correctly modeling propensity scores is crucial and can be challenging when dealing with high-dimensional confounders or non-linear effects. \tred{On the other hand, \citet{bang2005doubly} used a g-computation formula to estimate the longitudinal causal effect. However, accurately estimating the longitudinal outcome model and making a valid statistical inference when using machine learning are also crucial.}
To address these challenges, Doubly Robust (DR) methods, including augmented inverse probability weighting and targeted maximum likelihood estimation, have been introduced \citep{van2011targeted, kreif2017estimating, babino2019multiple}. DR estimators are derived by subtracting the projection onto Hilbert space and generally involve two nuisance parameters: {treatment estimator (e.g., propensity scores) and outcome estimator (e.g., iterative conditional mean)} \citep{yu2006double}. They are `doubly robust' in the sense that if either the treatment model or the outcome model is correct, the final estimate will be consistent. \tred{A widely used doubly robust estimation method for longitudinal data is longitudinal targeted maximum likelihood estimation (lTMLE; \citealp{gruber2010targeted,lendle2017ltmle}).}

More recently, machine learning-assisted causal inference methods have demonstrated effectiveness in handling high-dimensional covariates with complex, non-linear effects when estimating causal effects \citep{chernozhukov2018double, kunzel2019metalearners, chen2023effect, yang2023elastic}. However, most of these approaches are designed for studies with a single time point and are not well-suited for analyzing time-varying exposures in longitudinal settings. Notably, \cite{bodory2022evaluating} developed a double machine learning method and provided theoretical foundation to evaluate dynamic treatment effects. This approach innovatively integrates doubly robust estimation with machine learning algorithms, providing a data-driven solution to address time-varying confounding in longitudinal analyses. The underlying rationale for using machine learning in this context is that a doubly robust estimator achieves faster convergence rates when all nuisance functions are consistently estimated. This property enables the application of machine learning in causal settings, as the estimation bias and uncertainty from machine learning becomes negligible under mild conditions \citep{bodory2022evaluating, hernan2010causal, chernozhukov2018double}. Despite these advancements, the developed method remains sensitive to the selection of a single machine learning algorithm. In practice, more complex algorithms do not always yield optimal results \cite{chen2023effect}. One potential solution to address this limitation is the adoption of ensemble learning \citep{van2007super, chen2023effect}, which has gained popularity for its robust and superior performance achieved by integrating multiple machine learning algorithms. However, a comprehensive approach that combines causal machine learning with ensemble learning in longitudinal settings remains unexplored.





To bridge the gap, we propose a \textit{Marginal Structure Ensemble Learning Model} (\textbf{MASE}). \textbf{MASE} is developed based on the MSM solved by an efficient score, with the estimation of propensity scores and iterative conditional means as nuisance parameters. There are two major contributions in this paper: (1) from methodology perspective, we develop and evaluate a computationally efficient ensemble learning algorithm to integrate multiple machine learning algorithms to model nuisance functions, including time-varying propensity scores and a sequence of iterative outcome mean functions. We also evaluate the Neyman orthogonality condition and provide the analytical solution for the inference of estimated causal parameters. In numerical evaluations, we observe that our method effectively handles high-dimensional confounders, appropriately accounts for the longitudinal data structure, and outperforms existing methods in terms of estimation bias. (2) From an application perspective, \textbf{MASE} contributes to bridging gaps in our understanding of the long-term relationship between insufficient sleep and cognitive outcomes using ABCD data, compared to existing methods. Our findings indicate that sleep insufficiency can lead to cumulative declines in cognitive performance, providing complementary evidence to existing conclusions \citep{beebe2011cognitive}.

The rest of the paper is organized as follows. Section~\ref{sec:method} describes the proposed model, including the estimation process and inference method. Section~\ref{sec:data} showcases the application of our method to the ABCD study, outlining the effect estimation and presenting the inference results. Section~\ref{sec:simu} implements simulations to demonstrate the empirical performance of our method with finite sample sizes. We conclude the paper with a discussion in Section~\ref{sec:discuss}. Technical details and extra application results can be found in the Supplementary Material.

\section{Methods}\label{sec:method}
\subsection{Marginal Structural Model with Time-varying Exposure}\label{sec:MSM}
We consider a longitudinal study featuring a single binary exposure, multiple continuous outcomes, and high-dimensional confounders encompassing various data types. For $i = 1,\dots, n$, let $\mathcal{D}_i = (A_{it},\textbf{Z}_{it},\textbf{Y}_{it})$ represent the independent identically distributed data for subject $i$ across visit $t = 1,\dots,T$. Specifically, $A_{it} = a_{it}\in\{0,1\}$ denotes a univariate binary exposure variable, $\textbf{Z}_{it} = (Z_{it}^1,\dots,Z_{it}^{p_t})^T$ is a $p_t$-dimensional vector of observed confounders, with the number of covariates not necessarily being uniform at each visit. Further, $\textbf{Y}_{it} = (Y_{it}^1,\dots,Y_{it}^q)^T$ represents $q$ outcomes, which can be either continuous or categorical. For any variable $\{v_t\}_{1\le t\le T}$, we denote $\bar{v}_t = (v_1,\dots,v_t), t = 1,\dots,T$ as the historical measurements of variable $v$ up to the $t$-th visit. We define the counterfactual outcome \citep{rubin2005causal} for the $j$-th outcome at time $T$ as ${Y}_i^j(\bar{A}_{iT} = \{a_{i1},\dots,a_{iT}\})$ for subject $i$. The population mean of the counterfactual outcome is given by $\mathrm{E}({\textbf{Y}^j}(\bar{\textbf{A}}_T = \{\textbf{a}_1,\dots,\textbf{a}_T\})|{{\textit{\textbf{Z}}_0)}}$, where $\textit{\textbf{Z}}_0\in\mathbb{R}^{n\times p_0}$ is the baseline covariates matrix, and $\textbf{A}_t\in\mathbb{R}^n$ is the exposure vector.

\begin{figure}[!htbp]
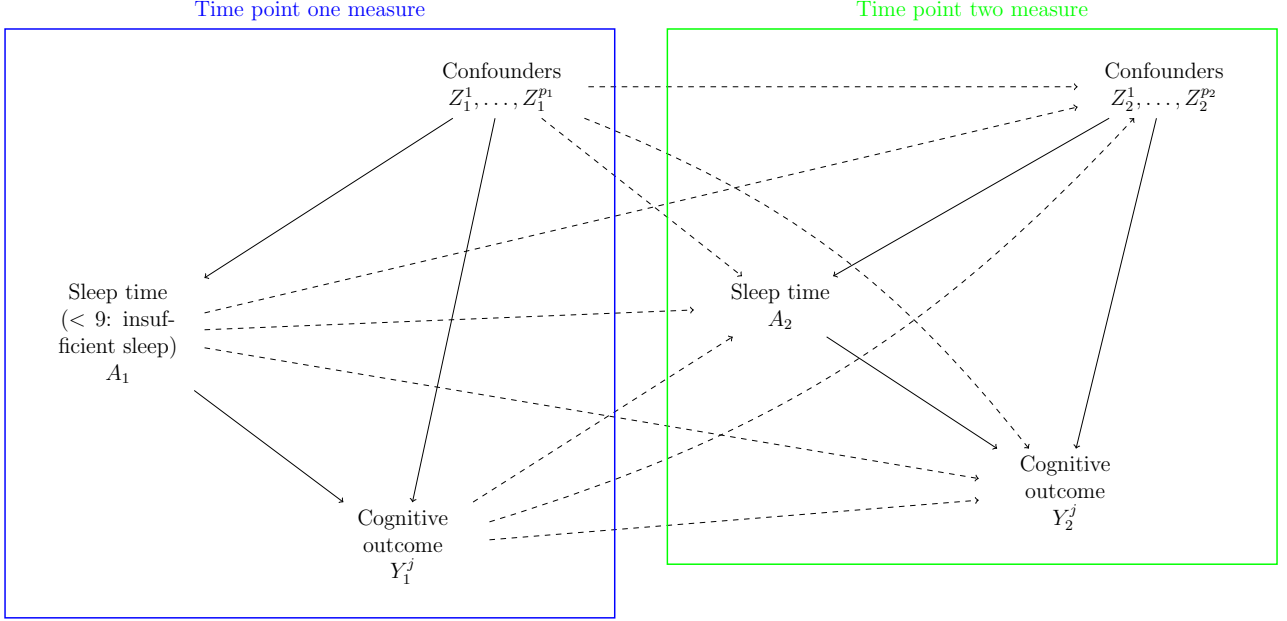

    \centering
    \scalebox{0.7}{
\centering
\tikz{
    \node (a1) [align=center, text width=3cm] {Sleep time \\($<9$: insufficient sleep)\\$A_1$};
    \node (z1) [above right=3cm and 4cm of a1, align=center,text width=3cm] {Confounders \\$Z_1^1,\dots,Z_1^{p_1}$};
    \node (a2) [below right=3cm and 2cm  of z1,align=center, text width=3cm] {Sleep time \\$A_2$};
    \node (z2) [above right=3cm and 4cm of a2,align=center, text width=3cm] {Confounders \\$Z_2^1,\dots,Z_2^{p_2}$};
    \node (y1) [below right= 3cm of a1,align=center, text width=3cm] {Cognitive outcome \\$Y_1^j$}; 
    \node (y2) [below right= 3cm of a2,align=center, text width=3cm] {Cognitive outcome \\$Y_2^j$}; 
    \path[->] (z1) edge (a1);
    \path[->] (z1) edge (y1);
    \path[->] (a1) edge (y1);
    \path[->] (z2) edge (a2);
    \path[->] (z2) edge (y2);
    \path[->] (a2) edge (y2);
    \draw[->,dashed] (a1) edge (a2);
    \draw[->,dashed] (a1) edge (z2);
    \draw[->,dashed] (a1) edge (y2);
    \draw[->,dashed] (z1) edge (a2);
    \draw[->,dashed] (z1) edge (z2);
    \draw[->,dashed] (z1) to[bend left=15] (y2);
    \draw[->,dashed] (y1) edge (a2);
    \draw[->,dashed] (y1) to[bend right=15] (z2);
    \draw[->,dashed] (y1) edge (y2);

    \node[draw=blue, thick, fit={(a1) (z1) (y1)}, inner sep=0.5cm, label=above:{\textcolor{blue}{Time point one measure}}] (group1) {};
    \node[draw=green, thick, fit={(a2) (z2) (y2)}, inner sep=0.5cm, label=above:{\textcolor{green}{Time point two measure}}] (group2) {};
}
}
    \caption{ABCD tabulate data structure. The solid lines represent the impact of variables to the others, the dashed line represents possible impact of previous measurements on the later ones. Notations $\textbf{A},\textbf{Z}, \text{ and } \textbf{Y}$ are defined in section~\ref{sec:MSM}.}
    \label{fig:intro}
\end{figure}

Figure~\ref{fig:intro} illustrates the directed acyclic graph (DAG) of our model, with all variables repeatedly measured at two time points (using a single outcome for demonstration purposes). Because both high-dimensional confounders and exposures are longitudinally measured and time-varying, estimating the aggregated causal effect is challenging. Classic methods, such as MSM, estimate the causal effect by leveraging propensity score weighting. Inspired by our data analysis, which suggests that previous cognitive outcomes can affect subsequent results (e.g., the crystallized composite score), we propose the following propensity score model that incorporates the effects of all time-varying confounders along with past outcomes:

\begin{gather}
        \pi_{i1}({a}_{i1},\textbf{z}_{i1}) = P({A}_{i1} = {a}_{i1}|\textbf{Z}_{i1} = \textbf{z}_{i1})\nonumber\\
        \pi_{it}(\bar{{a}}_{it},\bar{\textbf{z}}_{it},\bar{\textbf{y}}_{i(t-1)}) = P({A}_{it} = {a}_{it}|\bar{{A}}_{i(t-1)} = \bar{{a}}_{i(t-1)},\bar{\textbf{Z}}_{it} = \bar{\textbf{z}}_{it},\bar{\textbf{Y}}_{i(t-1)} = \bar{\textbf{y}}_{i(t-1)}), 2\le t\le T.
    \label{eq:PS}
\end{gather}
{
To ease the notation, we redefine $\bar{\textbf{Z}}_{it} = (\bar{\textbf{Z}}_{it},\bar{\textbf{Y}}_{i(t-1}), t = 2,\dots,T$. Then, the second PS function in~\eqref{eq:PS} can be simplified as $\pi_{it}(\bar{{a}}_{it},\bar{\textbf{z}}_{it})$.}

To ensure the identifiability of average causal effect $\tau(\bar{\textbf{a}}_t)$, we adopt the commonly used assumptions consistent with MSM. Specifically, in addition to the Stable Unit Treatment Value Assumption (SUTVA, \citealp{rubin1980randomization}), we assume that, for each individual $i$ and each  outcome $j$, the following hold: (1) consistency: ${{Y}_i^j}(\bar{{a}}_{iT}) = {Y}^j_{iT}$ if $\bar{{A}}_{iT} = \bar{{a}}_{iT}$; (2) no unmeasured confounding (NUC): for all $\bar{{a}}_{iT}$ and $t$, ${{Y}^j_i}(\bar{{a}}_{iT})\indep {A}_{it}|\overline{\textbf{{Z}}}_{it},\bar{{A}}_{i(t-1)} = \bar{{a}}_{i(t-1)}$; and (3) positivity: for all $t$ and $\bar{{a}}_{it}$, $\pi_{it}(\bar{{a}}_{it},\bar{\textbf{z}}_{it})>0$.
Throughout the paper, we refer to these assumptions as the Multivariate Longitudinal Sequential (MLS) assumptions. We then provide the MSM for the $j$-th outcome as:
\begin{equation}
    \mathrm{E}(\textbf{Y}^j(\bar{\textbf{a}}_T){{|\textbf{Z}_0}}) = \tau(\bar{\textbf{a}}_T,{{\textbf{Z}_0}}; \mathbf{\theta}^j), \forall j,
    \label{eq:MSM}
\end{equation}
where $\tau(\cdot,\cdot;\cdot)$ is a known function that models the $j$-th outcome based on exposures and covariates at baseline, and $\mathbf{\theta}^j$ is the vector of estimands. For example, we have $\tau(\bar{\textbf{a}}_2,{{\textbf{Z}_0}};\mathbf{\theta}^j) = \beta^j_0+\beta^j_1\textbf{a}_1+\beta^j_2\textbf{a}_2$ as a linear function with $\mathbf{\theta}^j = (\beta^j_0,\beta^j_1,\beta^j_2)'$, where $\textbf{a}_1,\textbf{a}_2$ correspond to the adolescent sleep time measured at visits one and two. Moreover, with effect modification, we can have $\tau(\bar{\textbf{a}}_2,{{\textbf{Z}_0}};\mathbf{\theta}^j) = \beta^j_0+\beta_{z0}^j\textbf{z}_0+\beta^j_1\textbf{a}_1+\beta^j_2\textbf{a}_2+\beta_{z1}^j\textbf{z}_0*\textbf{a}_1+\beta_{z2}^j\textbf{z}_0*\textbf{a}_2$. We will focus on the previous model with a simple linear form, {and we are interested in estimating the average treatment effect (ATE) for population getting sufficient sleep all the time (treatment regime $(0,0)$) versus population not having sufficient sleep at all (treatment regime $(1,1)$) as: $ATE := {\theta}^j_{ATE} =  \beta_1^j+\beta_2^j, \forall j$.}

\subsection{Doubly Robust Estimator}\label{sec:DR}

{Propensity score (PS) weighting is a common approach for estimating  the MSM equation~\eqref{eq:MSM}. The weights can be calculated as ${w}_{it} = {1}/{\Pi_{t = 1}^T\pi_{it}(\bar{{a}}_{it},\bar{\textbf{z}}_{it})}$. However, when massive time-varying confounders are present with potentially non-lnear confounding effects, a PS model based on a single machine learning algorithm could be incorrect due to numerical challenges. These incorrect modeling can lead to inconsistent estimates of causal effects within the MSM framework. Therefore, we are motivated from the doubly robust estimator \citep{babino2019multiple} and aim to develop a causal machine and ensemble learning-based estimator to mitigate the issue of model mis-specification and machine learning algorithm selection. }

{ 
The doubly robust estimator combines the estimates of conditional means and propensity score weights, ensuring consistent estimates of causal effects within the MSM framework when either the estimate of conditional means or propensity scores is corretly specified.  We first introduce an Iterative Conditional Mean Estimation (ICE) procedure: }

\begin{gather}
    \eta^j_T(\bar{\textbf{a}}_T,\bar{\textbf{z}}_T) = \mathrm{E}(\textbf{Y}_T^j|\bar{\textbf{A}}_T = \bar{\textbf{a}}_T, \bar{\textbf{Z}}_T = \bar{\textbf{z}}_T)\nonumber\\
    \eta^j_t(\bar{\textbf{a}}_T,\bar{\textbf{z}}_t) = \mathrm{E}(\eta^j_{t+1}(\bar{\textbf{a}}_T,\bar{\textbf{Z}}_{t+1})|\bar{\textbf{A}}_t = \bar{\textbf{a}}_t,\bar{\textbf{Z}}_t = \bar{\textbf{z}}_t), t = T-1,T-2,\dots,1\nonumber\\
    \eta^j_0(\bar{\textbf{a}}_T{{,\textbf{z}_0}}) = \mathrm{E}(\eta^j_1(\bar{\textbf{a}}_T,\textbf{Z}_1){{|\textbf{Z}_0 = \textbf{z}_0}}),\label{eq:ICE}
\end{gather}
for any $j = 1,\dots,q$. 

For simplicity, in the rest of this paper, we will eliminate the superscript $j$ for $\eta$'s.
{The doubly robust estimator for MSM is derived from the following estimating equation. Without loss of generality, we demonstrate the estimator with two timepoints. The estimating equation integrates the information from both conditional means and propensity scores \citep{babino2019multiple}:  }
\begin{equation}
    E(\textbf{S}_0+\textbf{S}_1+\textbf{S}_2) = \mathbf{0},
    \label{eq:DREE}
\end{equation}
where:
\begin{gather}
        \textbf{S}_{2} = \frac{d(\bar{\textbf{A}}_{2},\textbf{Z}_{0})}{\pi_1(\textbf{A}_{1},\textbf{Z}_{1})\pi_2(\bar{\textbf{A}}_{2},\bar{\textbf{Z}}_{2})}\{\textbf{Y}_{2}-\eta_2(\bar{\textbf{A}}_{2},\bar{\textbf{Z}}_{2})\}\nonumber\\
        \label{eq:DR}\textbf{S}_{1} = \sum_{\textbf{a}_{2}\in \mathbb{A}_{2}}\frac{d(\textbf{A}_{1},\textbf{a}_{2},\textbf{Z}_{0})}{\pi_1(\textbf{A}_{1},\textbf{Z}_{1})}\{\eta_2(\textbf{A}_{1},\textbf{a}_{2},\bar{\textbf{Z}}_{2})-\eta_1(\textbf{A}_{1},\textbf{a}_{2},\textbf{Z}_{1})\}\\
        \textbf{S}_{0} = \sum_{\textbf{a}_{1}\in\mathbb{{A}}_{1}}\sum_{\textbf{a}_{2}\in\mathbb{A}_{2}}d(\bar{\textbf{a}}_{2},\textbf{Z}_{0})\{\eta_1(\bar{\textbf{a}}_{2},\textbf{Z}_{1})-\tau(\bar{\textbf{a}}_{2}, \textbf{Z}_{0}; \mathbf{\theta}^j)\}.\nonumber
\end{gather}
Here, $\tau(\bar{\textbf{a}}_2, \textbf{Z}_{0}; \mathbf{\theta}^j)$ is given in~\eqref{eq:MSM}, $\mathbb{A}_t, t = 1,2$ represents the exposure set, for example, $\mathbb{A}_2 = \{\textbf{a}_{1} = \{\mathbf{0},\mathbf{1}\},\textbf{a}_{2} = \{\mathbf{0},\mathbf{1}\}\}$. 
Additionally, $d(\cdot)$ is any function with the same dimension as the parameter of interest $\mathbf{\theta}^j$. A simple example is to take $d(\cdot) = {\partial\tau(\bar{\textbf{a}}_2{{,\textbf{Z}_0}};\mathbf{\theta}^j)}/{\partial\mathbf{\theta}^j}$. When $\tau(\cdot,\cdot;\mathbf{\theta}^j)$ is a linear function, a closed-form solution for the estimating equation is available (see Supplementary Section~A) . 
For a non-linear $\tau(\cdot,\cdot;\mathbf{\theta}^j)$, we can solve equation~\eqref{eq:DREE} using the Newton-Raphson algorithm.

{In equation~\eqref{eq:DREE}, the estimate $\hat{\mathbf{\theta}}^j$ of $\mathbf{\theta}^j$ is consistent if either the estimates of $\pi_1,\pi_2$ or $\eta_1,\eta_2$ are correctly specified. {Further, equation~\eqref{eq:DREE} satisfies the Neyman Orthogonality condition \citep{neyman1959optimal}, which ensures fast convergence rate under correctly specified models for treatments (PS) and outcomes (ICEs) and justifies the use of machine learning methods in causal inference \citep{chernozhukov2018double}. We provide the formal theoretical definition in Lemma~\ref{lemma}.} }


\subsection{Ensemble Learning For time-varying PS and ICEs}
 Due to the high dimensionality of the time-varying confounders, which may have potential complex confounding effects in practical applications, the models for estimating both conditional means and propensity scores could be both incorrect if the machine learning algorithm is arbitrarily chosen. Since various machine learning algorithms can yield different performances under diverse scenarios, ensemble learning offers a more robust and generally more accurate solution. Besides, unlike the propensity score model, a compatible parametric model for iterative conditional mean may not be available in MSM under a generic setting \citep{babino2019multiple}. {In response to these challenges, we introduce a more flexible and computationally powerful ensemble learning-based procedure for the estimation of propensity scores and ICEs.}  

\textit{Ensemble procedure}: {Ensemble learning is a technique that combines several models to achieve improved generalization performance \citep{ganaie2022ensemble}. Let $M_1$ denote a set of candidate machine learning  algorithms (e.g., logistic regression, random forest, and gradient boosting) used in the ensemble learning procedure. The estimates of propensity scores for the $M_1$ candidate algorithms are $\hat{\pi}_t^{(1)},\dots,\hat{\pi}_t^{(M_1)}$ , where  $t = 1,\dots,T$ denotes the timepoints. Similarly, we define $\hat{\eta}_t^{(1)},\dots,\hat{\eta}_t^{(M_2)}$ as the fitted models of the ICEs for $M_2$ candidate algorithms. Given all fitted models (for details, see the estimation procedure in \textit{Model Fitting}), we aggregate them using the following weighted estimators:}
\begin{gather}
    \hat{\pi}_{EnL,t}(\bar{\textbf{a}}_t) = {g^{-1}_{EnL,t}}(\sum_{m_1 = 1}^{M_1}\hat{\omega}_{m_1}^{(1)}\hat{\pi}_t^{(m_1)}(\bar{\textbf{a}}_t))\nonumber\\
    \hat{\eta}_{EnL,t}(\bar{\textbf{a}}_t) = \sum_{m_2 = 1}^{M_2}\hat{\omega}_{m_2}^{(2)}\hat{\eta}_t^{(m_2)}(\bar{\textbf{a}}_t),
\label{eq:EnL}
\end{gather}
where $\hat{\mathbf{\omega}}^{(1)}$ and $\hat{\mathbf{\omega}}^{(2)}$ are the weights of the ensemble model for the propensity scores and iterative conditional means, respectively. {The function $g^{-1}_{EnL,t}(\cdot)$ bounds the weighted sum between $[0,1]$ (e.g., using the logit function)}. The weights can be determined by minimizing selected metrics that quantify the dissimilarity between the estimates and the data. For example, we can use the coefficients from logistic and linear regressions as weights to combine the candidate estimates.

We further adopt a cross-fitting approach to avoid the potential overfitting issue associated with the ensemble learning procedure \citep{chernozhukov2018double,chen2023effect}. Firstly, we split the data into training and testing datasets. {Throughout the model building procedure, we  train the base models using training data while constructing the ensemble models by leveraging testing data. We then repeat the model building procedure by exchanging the roles of training and testing datasets. Last, we estimate the parameter of interest (ATE) by solving:
\begin{equation}\label{eq:CFest}
    [E_{ts}(\Psi(\theta,\hat{\eta}_{tr},\hat{\pi}_{tr})+E_{tr}(\Psi(\theta,\hat{\eta}_{ts},\hat{\pi}_{ts})] = \mathbf{0},
\end{equation}
where $\Psi(\theta,\eta,\pi) = \textbf{S}_0+\textbf{S}_1+\textbf{S}_2$ is the score function as defined in~\eqref{eq:DREE}. $\hat{\eta}_{tr},\hat{\pi}_{tr}$ and $\hat{\eta}_{ts},\hat{\pi}_{ts}$  are nuisance parameter models estimated based upon training and testing data, respectively. $E_{ts},E_{tr}$ denote the expectations for testing and training data. 
}

\begin{figure}[!htbp]
    \centering
    \includegraphics[height=11cm,width=17cm]{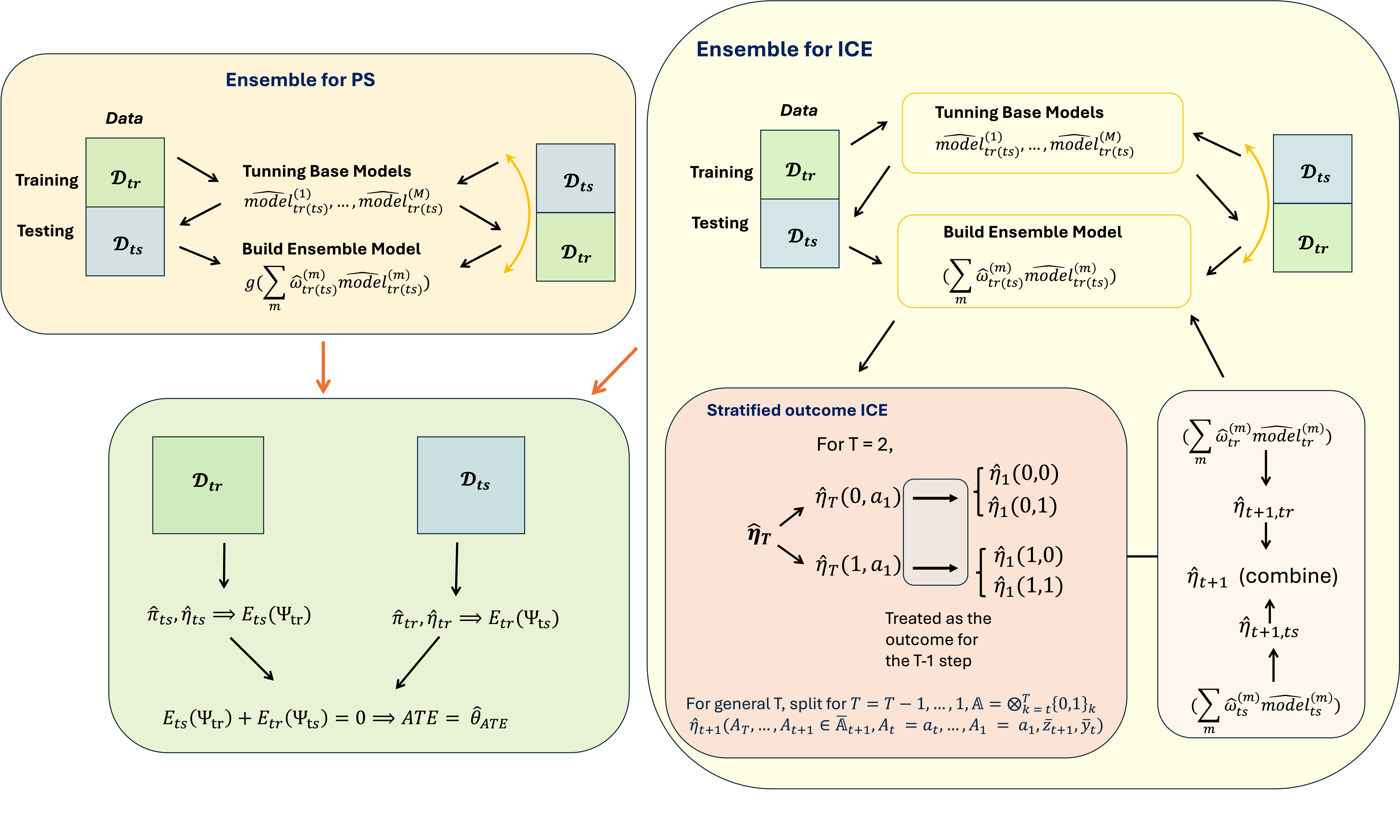}
    \caption{Overview of \textbf{MASE}: the left top box is the procedure for estimating propensity scores (PS), the right box is the procedure for estimating the outcome regressions (ICE); The left bottom box it the illustration of estimating the ATE ($\hat{\theta}$). We illustrate the stratification estimate for ICE within the right box (orange box).}
    \label{fig:DRMASE}
\end{figure}

\textit{Model fitting:} We remark here that learning iterative conditional means requires non-trivial effort. Unlike propensity scores, where compatible models can be specified by T models as in~\eqref{eq:PS} over the time course, the 
{ICE procedure relies on a sequence of predicted values of $\{\eta_t \}_{t=1}^T$. Identifying a unified compatible model for this sequence is not straightforward in general  \citep{babino2019multiple}. To alleviate this issue, we further extend the ensemble learning approach to facilitate sequential predictions, without imposing a unified model across the time course. As illustrated in ~\eqref{eq:ICE}, our goal is to estimate $ \eta_t (\bar{\textbf{a}}_T, \bar{\textbf{z}}_t), t = 1,\dots,T$ within a specific combination of treatments $\bar{\textbf{a}}_T$ with:
\begin{equation}\label{eq:ICE_EnL}
    {\eta}_{t}(\{\textbf{A}_T,\dots,\textbf{A}_{t}\}\in\bar{\mathbb{A}}_t,\textbf{A}_{t-1} = \textbf{a}_{t-1},\dots,\textbf{A}_1 = \textbf{a}_1,\bar{\textbf{z}}_t),
\end{equation}
where $\bar{\mathbb{A}}_t = \bigotimes_{k = t}^{T}\{0,1\}_k$ represents all possible combination of $\{0,1\}$ treatment regimes from timepoint $t$ to $T$. Consequently, we have $2^{T-t+1}$ predictions for ~\eqref{eq:ICE_EnL}. {For the cross-fitting procedure of $\eta$'s, we estimate the sequences of~\eqref{eq:ICE_EnL} on training and testing separately. Specifically, for each time point, we construct the models for $\eta$'s based on training data, and then apply the estimated $\eta$'s to predict~\eqref{eq:ICE_EnL} for testing data. We then repeat this procedure by exchanging training and testing dataset and iteratively construct $\eta$'s as in~\eqref{eq:ICE}. 
In each iteration, we predict $\eta_{t+1}(\{\textbf{A}_T,\dots,\textbf{A}_{t+1}\}\in\bar{\mathbb{A}}_{t+1},\textbf{A}_{t} = \textbf{a}_{t},\dots,\textbf{A}_1 = \textbf{a}_1,\bar{\textbf{z}}_{t+1})$ as the outcome for the next iteration to estimate $\eta_t(\bar{\textbf{a}}_T, \bar{\textbf{z}}_t)$.
The nuisance parameters related to $\eta$'s in~\eqref{eq:DR} can be straightforwardly estimated using $\hat{\eta}_t (\cdot)$ at each timepoint $t$.
Lastly, we follow the aforementioned cross-fitting procedure to estimate the ATE's. We summarize the above procedure in Algorithm~\ref{alg:MSMDR} with illustration in Figure~\ref{fig:DRMASE}.} \tred{The split of counterfactual outcomes distinguishes our method with Super Learner method \citep{van2007super} and aligns with the theoretical results in \citet{bodory2022evaluating}. }

}

\begin{algorithm}
\caption{\textbf{MASE}}\label{alg:MSMDR}
\begin{algorithmic}
    \State Input data $\mathcal{D}$
    \State Split data into $\mathcal{D}_{tr},\mathcal{D}_{ts}$
    \State \textit{Propensity Score Step}
   \For{t from T to 1}
   \State Estimate $\pi_t$ in~\eqref{eq:PS} by:
   \State (1) Tune the candidate base models $\hat{\pi}_t^{(m_1)},m_1 = 1,\dots,M_1$ using $\mathcal{D}_{tr}$
   \State (2) Build ensemble model $\hat{\pi}_{EnL,t,tr}$ as in~\eqref{eq:EnL} using $\mathcal{D}_{ts}$
   \State (3) Swap $\mathcal{D}_{ts}$ as training and $\mathcal{D}_{tr}$ as testing and repeat $(1)-(2)$, let $\hat{\pi}_{EnL,t,ts}$ be the trained ensemble model
   \EndFor 
   \State \textit{ICE Step}
   \State For $\forall j = 1,\dots,q$:
   \State Estimate $\eta_T$ with the similar procedure as in \textit{Propensity Score Step}, define $\hat{\eta}_{EnL,T,tr}$ as the ensemble model built by using $\mathcal{D}_{tr}$ as training; $\hat{\eta}_{EnL,T,ts}$ as the ensemble model built by using $\mathcal{D}_{ts}$ as training

   \For{t from (T-1) to 1}
       \State (I) For training data $\mathcal{D}_{tr}$, predict
       $\hat{\eta}_{EnL,t+1,ts}(\{\textbf{A}_T,\dots,\textbf{A}_{t+1}\}\in\bar{\mathbb{A}}_{t+1},\textbf{a}_{t},\dots,\textbf{a}_1, \bar{\textbf{z}}_{t+1})$
        \State (II) For testing data $\mathcal{D}_{ts}$, predict $\hat{\eta}_{EnL,t+1,tr}(\{\textbf{A}_T,\dots,\textbf{A}_{t+1}\}\in\bar{\mathbb{A}}_{t+1},\textbf{a}_{t},\dots,\textbf{a}_1, \bar{\textbf{z}}_{t+1})$ 
        \State (III) Let $\hat{\eta}_{EnL,t+1} = \{\hat{\eta}_{EnL,t+1,ts},\hat{\eta}_{EnL,t+1,tr}\}$ be the outcomes, build $\hat{\eta}_{EnL,t,tr}$ with~\eqref{eq:EnL} with $\mathcal{D}_{tr}$ as training; and $\hat{\eta}_{EnL,t,ts}$ with $\mathcal{D}_{ts}$ as training

    \EndFor
    \State Solve equation~\eqref{eq:CFest} using the models fitted above for $\hat{\theta}_{ATE}$\\
   
    \Return $\hat{\theta}_{ATE}$
\end{algorithmic}
\end{algorithm}

\textit{Theoretical results}: 
In Lemma~\ref{lemma}, {we show that our estimating equation satisfy the} Neyman Orthogonality condition \citep{neyman1959optimal}. This lemma implies that our proposed framework is robust to nuisance parameter estimation, and validates the use of machine learning in our framework {with valid inference results \citep{chernozhukov2017double}. The proof is provided in Supplementary material Section~A.}
\begin{lemma}\label{lemma}
    Let $\eta_t^*,\pi_t^*$ be the true values of $\eta_t,\pi_t,t = 1,2$. Suppose $T$ is a convex set such that $(\eta,\pi) \in T$ and $\mathcal{T}_N$ is a nuisance realization set which the estimators of nuisance parameters take values in this set with high probability. Assume the score function $\Psi(\theta,\eta,\pi) = \textbf{S}_0+\textbf{S}_1+\textbf{S}_2$ is measureable with Borel $\sigma$-fields for $\theta$ and $(\eta,\pi)$. Then $\Psi(\theta,\eta,\pi)$ satisfies: \\
        (i)\quad $\mathrm{E}(\Psi(\theta,\eta^*,\pi^*)) = 0$\\
        (ii)\quad $\frac{\partial}{\partial r}\mathrm{E}\{\Psi(\theta,(\eta^*,\pi^*)+r(\delta_{\eta},\delta_{\pi}))\}$ exists for all $r\in[0,1)$\\
        (iii)\quad $\frac{\partial}{\partial r}|_{r=0}\mathrm{E}\{\Psi(\theta,(\eta^*,\pi^*)+r(\delta_{\eta},\delta_{\pi}))\} = 0$,\\
    with respect to the nuisance realization set $\mathcal{T}_N\subset T$, where $\delta_{\eta} = \eta-\eta^*,\delta_{\pi} = \pi-\pi^*$. 
\end{lemma}

{Following \cite{chernozhukov2018double,chernozhukov2017double}, we derive the statistical inference procedure for \textbf{MASE}. We first calculate the plug-in estimator:}
\begin{gather}
        \mathbf{J}_n(\hat{\theta}) = \frac{1}{n}\sum_{i = 1}^n\{\frac{\partial}{\partial \theta}\Psi_i(\hat{\theta},\hat{\eta},\hat{\pi})\}\nonumber\\
    \mathbf{F}_n(\hat{\theta}) = \frac{1}{n}\sum_{i = 1}^n\{\Psi_i(\hat{\theta},\hat{\eta},\hat{\pi})\Psi_i(\hat{\theta},\hat{\eta},\hat{\pi})^T\}\nonumber\\
    \label{eq:var}\mathbf{V}_n(\hat{\theta}) = \mathbf{J}_n(\hat{\theta})^{-1}\mathbf{F}_n(\hat{\theta})[\mathbf{J}_n(\hat{\theta})^{-1}]^T.
\end{gather}
Then, the $(1-\alpha)\times 100$ confidence interval can be calculated by:
\begin{equation}
    CI := [\hat{\theta}_{ATE}\pm \Phi^{-1}(1-\alpha/2)\hat{\sigma}_{ATE}/\sqrt{n}],
\end{equation}
with $\hat{\sigma}_{ATE}^2 = \sum_{i = 2}^{T+1}\mathbf{V}_{n,ii}+2\sum_{i=2<j}\sum_{j = 2}^{T+1}\mathbf{V}_{n,ij}$, where $\mathbf{V}_{n,ij},i,j = 1,\dots,T+1$ are the $\{i,j\}$-th element in $V_n(\hat{\theta})$ matrix. $\hat{\eta}$ and $\hat{\pi}$ are the ensemble learning estimated nuisance parameters from combining the estimates for training and testing datasets.

\section{Results}\label{sec:data}
{We applied \textbf{MASE} to the ABCD study to investigate the effect of sleep insufficiency on cognitive development in adolescents. The exposure variable was adolescent sleep insufficiency, defined as a binarized measure of sleeping time with a cutoff of nine hours. The outcomes were cognitive function scores measured at baseline and at two follow-up timepoints during the second and fourth-year visits. The age-corrected measures from the NIH cognitive toolbox were used, including the Picture Vocabulary Test (picvocab), Flanker Test (flanker), Pattern Comparison Processing Speed Test (pattern), Picture Sequence Memory Test (picture), Oral Reading Recognition Test (reading), and the Crystallized Composite Scores (cryst).

The ABCD data version $5.0$ initially recruited $11,868$ participants at the first visit, $10,973$ at the two-year follow-up, and $4,754$ at the four-year follow-up. {Our study includes $7,070$ participants with comprehensive cognitive function measures at the first two visits.}
We considered demographic variables including age, sex at birth, race, ethnicity, family income, the youth's education, parent education, a general latent factor of economic, social, and physiological well-being, a latent factor for youth perceived social support, and a latent factor for perinatal health. Additionally, we explored potential confounders from nutrition intake (breakfast and diet nutrition), neighborhood, parent behavior inventory, and genetic data (principle component of genetic ancestry). We selected the confounders using the sure screening method \citep{fan2008sure} and included those related to both exposures and outcomes. The final list of potential confounders included the seven demographic variables, $104$ nutrition variables from breakfast and eight variables from parent behavior inventory. 
The list of confounders and  baseline demographics  were summarized in Supplementary Section~B. 

We implemented the ensemble learning by cross-fitting using candidate algorithms included linear (logistic) regression, elastic net, XGboosting, and Random Forest. The meta learners (ensemble models) employed were simple linear (logistic) regressions. We utilized models from \textit{sklearn} library in Python $3.9$ and tuned basic parameters such as maximum depth, learning rate, and number of estimators for tree models (boosting), as well as penalty parameters for regression-based models (e.g., elastic net). \tred{To avoid potential positivity violation issue, we trimmed the propensity score to lie within the range of $0.01$ and $0.99$ \citep{schomaker2019using}.}
Furthermore, we compared \textbf{MASE} with {MSM with propensity score estimated by logistic regression (MSM-lm), ICE with linear regression (ICE-lm), and longitudinal targeted maximum likelihood combining with super learner (base learners are: linear (logistic) regression, random forest, XGboosting, and elastic net). The standard error estimates for MSM-lm and ICE-lm were calculated by bootstrap method, whereas those of \textbf{MASE} and lTMLE were from analytic solutions.}


We summarized the long-term effect results for the age-corrected outcomes in Figure~\ref{fig:NIHTBX}. The estimated values along with their confidence intervals are in Table~\ref{tab:NIHTBX}. We observed significantly negative effects of insufficient sleep on the cognitive development in several measurements. This included the Oral Reading Recognition Test ($-1.90[-3.41,-0.39]$), the Picture Vocabulary Test ($-2.81[-4.11,-1.51]$), the Pattern Comparison Processing Speed Test ($-2.40[-4.50,-0.30]$), and the Crystallized Composite Score ($-2.59[-3.98,-1.20]$). Notably, the Crystallized Composite Score, which is expected to develop significantly during adolescence \citep{heaton2014reliability}, showed substantial negative impacts. 

\begin{figure}[!htbp]
\centering
        {\includegraphics[width=12cm,height=12cm]{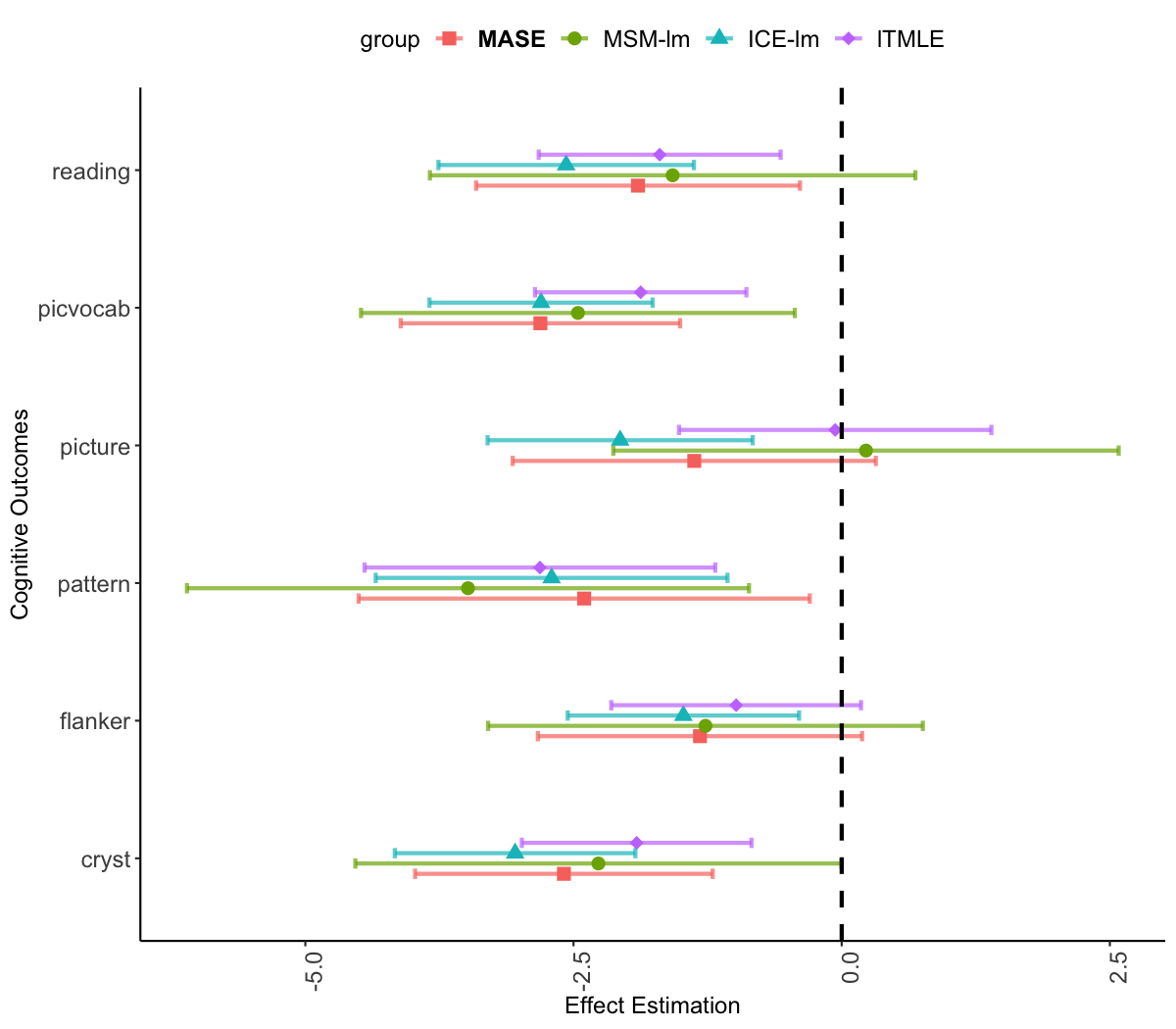}}

    \caption{Forest plots for long-term effects estimates of insufficient sleep on cognitive development on ABCD data. The reference level is sufficient sleep. Estimates from different methods are represented using different types of points. 
    }
    \label{fig:NIHTBX}
\end{figure}

\begin{table}[!htbp]
    \centering
    \scalebox{0.9}{
    \begin{tabular}{*{5}{c}}
    \hline\hline
    NIH Toolbox&\\
    & \textbf{MASE}& MSM-lm & ICE-lm&lTMLE\\
    \textbf{Age-corrected}\\
    reading&$-1.90$&$-1.58$&$-2.57$&$-1.70$\\
    &$[-3.41,-0.39]*$&$[-3.84,0.69]$&$[-3.76,-1.38]*$&$[-2.82,-0.57]*$\\
    picvocab&$-2.81$&$-2.46$&$-2.80$&$-1.87$\\
    &$[-4.11,-1.51]*$&$[-4.48,-0.44]*$&$[-3.84,-1,76]*$&$[-2.86,-0.89]*$\\
    picture&$-1.37$&$0.23$&$-2.07$&$-0.06$\\
    &$[-3.07,0.32]$&$[-2.13,2.58]$&$[-3.30,-0.83]*$&$[-1.52,1.40]$\\
    pattern&$-2.40$&$-3.48$&$-2.70$&$-2.81$\\
    &$[-4.50,-0.30]*$&$[-6.10,-0.86]*$&$[-4.34,-1.06]*$&$[-4.45,-1.18]*$\\
    flanker&$-1.32$&$-1.27$&$-1.48$&$-0.98$\\
    &$[-2.83,0.19]$&$[-3.30,0.76]$&$[-2.55,-0.40]*$&$[-2.15,0.18]$\\
    cryst&$-2.59$&$-2.27$&$-3.04$&$-1.91$\\
    &$[-3.98,-1.20]*$&$[-4.53,-0.002]*$&$[-4.17,-1.92]*$&$[-2.98,-0.84]*$\\

        \hline\hline
    \end{tabular}}
     \caption{Causal effect estimation (insufficient sleep vs. sufficient sleep) and 95\% confidence interval (in parentheses) for age-corrected outcomes in ABCD study. * represents the significant CIs.}
    \label{tab:NIHTBX}
\end{table}


In summary, our results demonstrate statistically significant (adverse) effects of long-term insufficient sleep on cognitive development. These findings provide insights for developing intervention strategies to address delayed cognitive development. {In comparison, MSM has wider confidence intervals. All the other methods performed relatively consistent.} {As part of the additional sensitivity analyses, we have included the results based on age-uncorrected results and the analysis for three time points measurements (for both age-corrected and age-uncorrected). We summarized such results in Supplementary Section~B.} We also checked the positivity assumption with propensity scores calculated using the ensemble methods, there is no severe violation (Supplementary B).

\section{Simulation}\label{sec:simu}
{We evaluated the performance of our method and compared it with comparable methods in simulation studies. We generated datasets with various numbers of time-varying confounders (e.g., 50 and 100) and different causal effect sizes. We considered a sample size of $n = 1,000$ to mimic our motivation datasets. Without loss of generality, we considered two timepoints $T=2$. Specifically, we simulated the data for $t = 1, 2$ by sampling: }
\begin{gather}
    \textbf{Z}_t\sim N(0,\mathbf{\Sigma}_t)\nonumber\\
    \pi_t|\bar{\textbf{Z}}_t = \{1+exp(\mu_{\pi_t}(\bar{\textbf{Z}}_t))\}^{-1}\nonumber\\
    \textbf{A}_t|\bar{\textbf{Z}}_t \sim \mathrm{Binomial}(\pi_t)\\
    \textbf{Y}_t|\bar{\textbf{Z}}_t,\bar{\textbf{A}}_t \sim N(\mu_t(\bar{\textbf{A}}_t,\bar{\textbf{Z}}_t),\mathbf{\Sigma}_Y),\nonumber
\end{gather}\label{eq:simu}
{\noindent where $\mathbf{\Sigma}_1,\mathbf{\Sigma}_2\in\mathbb{R}^{p\times p}$ were covariance matrices for confounders with pre-specified correlation structures, and $\mathbf{\Sigma}_Y$ was the covariance matrix for the outcome; $\mu_{\pi_1}(\textbf{Z}_1),\mu_{\pi_2}(\textbf{Z}_2)$ were the mean structures for propensity scores (line 2); and $\mu_1(\bar{\textbf{A}}_1,\bar{\textbf{Z}}_1),\mu_2(\bar{\textbf{A}}_2,\bar{\textbf{Z}}_2)$ were the mean structures for conditional means (line 4). We generated ${Y}_{it}$ based on the counterfactual outcomes sequence of the treatments form all previous time points $A_{it},A_{i(t-1)},\dots,A_{i1}$. For example, for $t = 2$, the outcome would be $Y_{i2} = A_{i1}A_{i2}Y(1,1)+A_{i1}(1-A_{i2})Y(1,0)+(1-A_{i1})A_{i2}Y(0,1)+(1-A_{i1})(1-A_{i2})Y(0,0)$. 
}

{To reflect the nonlinear effects of  the time-varying confounders, we specified $\mu_{\pi_t}(\textbf{Z}_t) = s_{\pi_t}(\bar{\textbf{Z}}_t,\gamma_{\pi_1}), t = 1,2$, where $s_{\pi_t}(\bar{\textbf{Z}}_t,\gamma_{\pi_t})$ was a nonlinear function. For example, $s_{\pi_t}$ included cosine functions of $\bar{\textbf{Z}}_t$ and their inner interactions. Similarly, we set nonlinear functions for $\mu_t(\bar{\textbf{A}}_t,\bar{\textbf{Z}}_t) = s_t(\bar{\textbf{A}}_t,\bar{\textbf{Z}}_t,\gamma_t), t = 1,2$ with a known parameter $\gamma_t$. We repeated each setting for 500 Monte Carlo simulations.}

\begin{table}[!htbp]
    \centering
    \scalebox{0.65}{
    \begin{tabular}{*{10}{c}}
    \hline\hline
   &\multicolumn{4}{c}{$p = 50$, Effect Size: $5$}&&\multicolumn{4}{c}{$p = 100$ Effect Size: $5$}\\
   &Estimation&Monte Carlo SD&Relative Bias&Estimated SE&&Estimation&Monte Carlo SD&Relative Bias&Estimated SE\\
   Monte Carlo&$5.024$&$0.049$&&&&$5.022$&$0.054$\\
   \textbf{MASE}&$4.821$&$0.427$&$\textbf{-0.040}$&$0.331$&&$5.221$&$0.455$&$\textbf{-0.040}$&$0.334$\\
   ICE-lm&$5.959$&$0.446$&$0.186$&$0.425$&&$5.906$&$0.457$&$0.176$&$0.455$\\
   MSM-lm&$5.673$&$0.589$&$0.129$&$0.736$&&$5.636$&$0.861$&$0.122$&$1.616$\\
   ltlme&$5.466$&$0.407$&$0.088$&$0.306$&&$5.491$&$0.406$&$0.093$&$0.220$\\
   \\
   &\multicolumn{4}{c}{$p = 50$, Effect Size: $3$}&&\multicolumn{4}{c}{$p = 100$ Effect Size: $3$}\\
   Monte Carlo&$2.999$&$0.048$&&&&$2.987$&$0.054$\\
   \textbf{MASE}&$3.199$&$0.428$&$\textbf{0.067}$&$0.374$&&$3.139$&$0.468$&$\textbf{0.051}$&$0.3349$\\
   ICE-lm&$3.926$&$0.439$&$0.309$&$0.419$&&$3.873$&$0.450$&$0.297$&$0.447$\\
   MSM-lm&$3.635$&$0.584$&$0.212$&$0.729$&&$3.597$&$0.818$&$0.204$&$1.585$\\
   lTMLE&$3.415$&$0.390$&$0.139$&$0.299$&&$3.437$&$0.390$&$0.151$&$0.211$\\
   \\
   &\multicolumn{4}{c}{$p = 50$, Effect Size: $1$}&&\multicolumn{4}{c}{$p = 100$ Effect Size: $1$}\\
   Monte Carlo&$1.152$&$0.048$&&&&$1.151$&$0.053$\\
   \textbf{MASE}&$1.364$&$0.397$&$\textbf{0.184}$&$0.341$&&$1.467$&$0.435$&$\textbf{0.275}$&$0.390$\\
   ICE-lm&$2.094$&$0.429$&$0.818$&$0.412$&&$2.036$&$0.440$&$0.769$&$0.440$\\
   MSM-lm&$1.798$&$0.572$&$0.561$&$0.719$&&$1.730$&$0.844$&$0.503$&$1.569$\\
   lTMLE&$1.589$&$0.337$&$0.379$&$0.293$&&$1.618$&$0.368$&$0.406$&$0.204$\\

        \hline\hline
    \end{tabular}
    }
     \caption{
     Results for $500$ Monte Carlo simulations. We compared the performance of all methods with sample size $n = 1000$,   different dimensions of confoudners ($p$), and various effect sizes. The estimated SE's are the averages of standard error estimates for the $500$ simulations.}
    \label{tab:simu}
\end{table}

We applied the \textbf{MASE} and various benchmark methods to the simulated datasets and summarized the results in Table~\ref{tab:simu}. The benchmark methods included the MSM with logistic estimated PS (MSM-lm), ICE with linear regression (ICE-lm), and commonly used longitudinal targeted maximum likelihood estimator along with super learner(lTMLE, \citealp{lendle2017ltmle,schomaker2019using}). {Specifically, we adopted multiple base learners for both \textbf{MASE} and lTMLE, including linear (logistic) regression, XGboosting, random forest, and elastic net.} The parameter of interest, $\mathbf{\theta}$, can be calculated by contrasting the expected values of counterfactual outcomes between treatments for each simulation iteration. 
We evaluated the performance using the estimated relative bias, which is calculated as ${(\hat{\mathbf{\theta}}-\mathbf{\theta})}/\mathbf{\theta}$, and compared the empirical standard deviation of $\hat{\mathbf{\theta}}$ with the {standard error estimated by \textbf{MASE}. For standard error estimation, analytical estimators were used for \textbf{MASE} and lTMLE, while bootstrap estimates were used for MSM-lm and ICE-lm. }


In general, \textbf{MASE} outperformed the competitive methods, exhibiting lower relative bias across all settings. {The analytic standard error estimates from our method approximated the Monte Carlo standard deviations, suggesting the validity of the proposed estimate.}

\section{Discussion}\label{sec:discuss}
{We have developed a new tool \textbf{MASE}, to estimate the causal effects of time-varying exposures on multiple outcomes in longitudinal studies in the presence of high-dimensional confounders with complicated confounding mechanism. {This method is tailored to handle numerous potential confounders with non-linear confounding effects by integrating various machine learning algorithms in estimating time-varying PS and ICEs. Additionally, we have explored its theoretical properties for validating the use of machine learning and provided a computationally efficient algorithm for implementing \textbf{MASE}. Simulation studies have demonstrated the validity of \textbf{MASE} in terms of less estimation bias compared to existing methods. \tred{The proposed time-varying ensemble and cross-fitting procedure in \textbf{MASE} addresses the compatibility issue in the estimation of sequential expectations, thereby improving performance compared to lTMLE.}}}

{In our ABCD analysis, the results revealed that sleep insufficiency can lead to aggregated declines in cognitive performances. These findings align more neurobiologically with existing literature (see \citealp{beebe2011cognitive}) compared to previous findings    (Figure~2 in \citealp{yang2022effects}), which suggested that sleep insufficiency `improves' pattern-related cognition by using cross-sectional causal inference tools in the longitudinal ABCD study.
{Our findings clarify the long-term causal effects of sleep insufficiency on specific cognitive functions, which may further guide effective prevention strategies (e.g., restricting electronic device usage time at night) for modifiable risk factors of jeopardized cognitive function development in youth. \tred{Further, our work can be extended to investigate the heterogeneity of treatment effects. A potential consideration is a second stage machine learning regression method on unentered influence function of potential outcomes \citep{takatsu2024doubly}.} The package for \textbf{MASE} is available at  \url{https://github.com/zhuivv/DR-MASE}.} 
    }

\section*{Acknowledgement}
The data used in the preparation of this article were obtained from the Adolescent Brain Cognitive Development (ABCD) Study, accessible through the NIMH Data Archive (NDA) at \url{https://abcdstudy.org}. Chen, C's work was supported by National Institutes of Health grants R01 AG089377,
P30 AG028747, and by the Johns Hopkins Institute for Clinical and Translational Research (ICTR) funded under grant 1UM1TR004926 from the National Center for Advancing Translational Sciences (NCATS) a component of the National Institutes of Health (NIH), and NIH Roadmap for Medical Research. The content of this paper is solely the responsibility of the authors and do not necessarily represent the official view of the Johns Hopkins ICTR, NCATS, or NIH.

\bibliographystyle{apalike} 
\bibliography{ref}

\newpage



\end{document}